\documentstyle[11pt,aaspp4,epsf]{article}
\newcommand{\gapr}{\raisebox{-.6ex}{\mbox{
$\stackrel{>}{\mbox{\scriptsize$\sim$}}\:$}}}
\newcommand{\lapr}{\raisebox{-.6ex}{\mbox{
$\stackrel{<}{\mbox{\scriptsize$\sim$}}\:$}}}
\def\df{\dot{f}}

\def\Ts{T_{\rm span}}
\def\Z{Z^2_1}
\def\fh{f_{\rm HRI}}
\def\fp{f_{\rm PSPC}}
\def\rosat{{\sl ROSAT\/}~}
\def\asca{{\sl ASCA\/}~}
\def\ns{{RX~J0822--4300\/}~}
\begin{document}
\lefthead{Pavlov, Zavlin, Tr\"umper}
\righthead{X-ray Pulsations from the Central Source in Pup A}
\title{X-ray Pulsations from the Central Source in Puppis A}
\author{G.~G. Pavlov}
\affil{The Pennsylvania State University, 525 Davey Lab,
University Park, PA 16802, USA; pavlov@astro.psu.edu}
\and
\author{V.~E. Zavlin and J.~Tr\"umper}
\affil{Max-Planck-Institut f\"ur extraterrestrische Physik, D-85740
Garching, Germany; zavlin@xray.mpe.mpg.de}
%%%%%%%%%%%%%%%%%%%%%%%%%%%%%%%%%%%%%%%%%%%%%%%%%%%%%%%%%%%%%%%%%%%%%%
\begin{abstract}
There are several supernova remnants 
which contain unresolved 
X-ray sources close to their centers, presumably radio-quiet
neutron stars. To prove that these objects are indeed neutron stars,
to understand the origin of their X-ray radiation, and to explain
why they are radio-quiet, one should know their periods and
period derivatives. 
We searched for pulsations of the X-ray flux from the radio-quiet
neutron star candidate RX~J0822--4300 near the center of the
Puppis A supernova remnant observed with the \rosat PSPC and HRI. 
A standard timing analysis of the separate
PSPC and HRI data sets
does not allow one to detect the periodicity unequivocally.
However, a thorough analysis of the two
observations separated by 4.56~yr
enabled us to find a statistically significant
period $P\simeq 75.3$~ms and its derivative 
$\dot{P}\simeq 1.49\times 10^{-13}$~s~s$^{-1}$.
The corresponding characteristic parameters of the neutron star,
age $\tau=P/(2\dot{P}) = 8.0$~kyr, magnetic field
$B=3.4\times 10^{12}$~G, and rotational energy loss
$\dot{E}=1.4\times 10^{37}$~erg~s$^{-1}$, are typical
for young radio pulsars. 
Since the X-ray radiation has a thermal-like spectrum, its pulsations
may be due to a nonuniform temperature distribution over the
neutron star surface caused by anisotropy of the heat conduction
in the strongly magnetized crust.
\end{abstract}
%%%%%%%%%%%%%%%%%%%%%%%%%%%%%%%%%%%%%%%%%%%%%%%%%%%%%%%%%%%%%%%%%%%%
%%%%%%%%%%%%%%%%%%%%%%%%%%%%%%%%%%%%%%%%%%%%%%%%%%%%%%%%%%%%%%%%%%%%%%
\section{Introduction}
Observations with the {\sl Einstein}, \rosat and \asca
missions have revealed several isolated, radio-quiet neutron star (NS)
candidates in supernova remnants (see 
Brazier \& Johnston 1998, for a recent review).
These objects are characterized by
a lack of observable radio emission, thermal-like X-ray spectra
with typical (blackbody) temperatures of $\sim (1-5)\times 10^6$~K,
and very high ratios, $\gapr 1000 $, of the X-ray to optical fluxes.
Their X-ray radiation has been interpreted as thermal radiation
from cooling NSs. The reason for the lack of observable
radio emission remains unclear. The simplest explanation is that
these NSs are ordinary radio pulsars whose rotational and magnetic
axes are unfavorably oriented so that the pulsar beam cannot
be seen at Earth. An alternative explanation is that  these objects
are not active radio pulsars 
due to, e.g., too low magnetic fields or too slow rotation. 
Another option is that these NSs
belong to the class of anomalous X-ray pulsars,
also radio-quiet, slowly rotating NSs which apparently have superstrong,
$\sim 10^{14}$--$10^{15}$~G, magnetic fields (see, e.g., Vasisht \&
Gotthelf 1997). To choose among these hypotheses, one 
should know the rotation period and its derivative,
which would allow one to estimate the object's age and magnetic field. 
However, many attempts to find periods in the X-ray radiation of
these sources 
have been unsuccessful so far,
which indicates that the pulsed fraction may be too low
to distinguish between a signature of the true
periodicity and a noise fluctuation in a relatively short observation
with a small number of photons collected.
On the other hand, increasing the observational time 
strongly increases the number of trial periods (and period
derivatives in the case of very long observations).
This not only makes the search prohibitively expensive,
but also increases the probability of obtaining large noise
fluctuations, which hampers detecting the true periodicity. 
However, if two or more timing observations of a source
have been carried out, well separated in time,
with good time resolution and
sufficient amounts of photons collected, one may use
additional constraints to discriminate between true
and false periodicity signatures in the separate timing data. 
We applied such an approach to searching for periodicity of the 
X-ray brightest, 
radio-quiet NS candidate RX~J0822--4300 in the supernova remnant (SNR)
Puppis A, making use of \rosat PSPC and HRI observations.

Puppis A is one of 
three known oxygen-rich SNRs in the
Galaxy. Its kinematic age, $3.7\pm 0.4$~kyr, was estimated
from the proper motion of fast-moving filaments in the northeast quadrant
of the SNR 
(Winkler et al.~1988).
Its distance, $d=2.2\pm 0.3$~kpc, was estimated from 
VLA observations in 
the $\lambda$21 cm hydrogen line in the direction of Puppis A 
(Reynoso et al.~1995). The central compact source
was seen in the {\sl Einstein} HRI images, located $\simeq 6'$
from the kinematical center of SNR expansion (Petre et al.~1982).
Petre, Becker \& Winkler (1996; hereafter PBW96)
analyzed \rosat HRI and PSPC observations (exposure times
4~ks and 6~ks, respectively) and suggested that
this source is an isolated NS born in the supernova explosion. It 
has no optical
counterpart to limiting magnitudes $B\gapr 25.0$, $R\gapr 23.6$,
which implies an X-ray/optical flux ratio $f_X/f_B \gapr 5000$.
No obvious radio counterpart was detected to a 3$\sigma$ upper limit 
of 0.75~mJy at 1.4 GHz. Fitting the PSPC spectrum of the source 
with the blackbody model gives a temperature of $0.28\pm 0.10$~keV,
somewhat higher than $0.10-0.18$~keV expected from 
standard NS cooling models. The corresponding blackbody radius
of $\sim 2$~km is smaller than the expected radius for a NS.
No evidence for pulsations with a pulsed fraction larger than
20\% was found.

New observations of Puppis A and its central source with the
\rosat and \asca have been performed recently.
We used these data 
for further investigations of the spectrum and the light curve of
the NS candidate. 
Results of the spectral investigation will be published
elsewhere (Zavlin, Pavlov, \& Tr\"umper 1998b). Briefly,
they show that the temperature and radius of the NS are
compatible with standard models if to assume that the NS
surface is covered with a hydrogen/helium atmosphere
(cf.~Zavlin, Pavlov, \& Tr\"umper 1998a). 
Fitting the observed spectra
with the NS atmosphere models indicates that the surface magnetic
field may be rather high, $\gapr 4\times 10^{12}$~G,
so that one can expect
that the NS magnetic poles are hotter than the equator (Greenstein
\& Hartke 1983),
which may lead to observable modulation of the thermal X-ray flux. 
%%%%%%%%%%%%%%%%%%%%%%%%%%%%%%%%%%%%%%%%%%%%%%%%%%%%%%%%%%%%%%%%%%%%%%%%%%
%%%%%%%%%%%%%%%%%%%%%%%%%%%%%%%%%%%%%%%%%%%%%%%%%%%%%%%%%%%%%%%%%%%%%%%%%%
\section{Observational Data and Timing Analysis}
Among several observations of Puppis A in the public archive 
at HEASARC/GSFC, 
two are suitable for searching for periodicity of the central
source. The first one 
was  carried out with the \rosat PSPC on 1991 April 16 (6.0~ks
useful exposure, 53~ks total time
span between the beginning and the end of the observation).
In that observation the NS was off-set by $\simeq 12\farcm 5$.
The background-subtracted
count rate is $0.237\pm 0.006$~s$^{-1}$
($0.296\pm 0.007$~s$^{-1}$ after applying the dead-time and vignetting
corrections). For the period search, we used 1368 counts extracted
in the 0.1--2.4~keV range 
from a $25''$ radius aperture.

Another useful observation
was carried out with the \rosat HRI on 1995 November 2--8
(30.7~ks useful exposure, $580$~ks total time span).
We evaluated the background-subtracted source count
rate, $0.070\pm 0.002$~s$^{-1}$
($0.080\pm 0.003$ after the dead-time correction).
An analysis of the radial distribution of the
source counts does not show significant deviations from
the HRI Point Spread Function for a point source.
To reduce the background contamination, we used 1968 counts 
collected from a $10''$ radius aperture centered at the point source 
($\alpha_{2000}=8^{\rm h} 21^{\rm m} 57\fs 5$, 
$\delta_{2000}=-43^\circ 00' 14\farcs 5$). 

The \rosat HRI observations of 1991 
and 1992 were too short (2.4 and 1.5~ks, respectively)
to collect enough point source photons.
\ns was also observed with the \asca GIS (1993 July),
but a too low time resolution, 65.2~ms.

To search for periodicity of RX~J0822--4300,
we converted the spacecraft clock (SCC) arrival
times of the HRI and PSPC photons 
to UTC times, and then to barycentric dynamical times (TDB).
The SCC-UTC conversion was performed
with a fourth order polynominal fit of the SCC calibration points to UTC.
The fit residuals 
were within a 3~ms range; this corresponds to a scatter in pulse
phase $<0.1$ for frequencies $\lapr 30$~Hz.

To perform a global periodicity search of sparse data,
tests based on $Z_m^2$ statistics ($m$ is the number of
harmonics included)
look most appropriate (Buccheri et~al.~1983; De Jager 1994).
Since the point source spectrum is likely of a thermal origin
(PBW96; Zavlin et al.~1998b), 
the pulse profile is expected to be smooth, with
main contributions coming from the first harmonic,
as observed in thermal X-ray radiation from radio pulsars
(e.g., Anderson et al.~1993) and predicted by NS atmosphere
models (Shibanov et al.~1995). This means 
that the $\Z$ (Rayleigh) test
should be optimal for the periodicity search:
the ephemeris parameters $f$ and $\dot{f}$ 
are estimated as the values that give
the largest value of 
$\Z=2N^{-1}[(\sum_{i=1}^N \cos2\pi\phi_i)^2
+(\sum_{i=1}^N \sin2\pi\phi_i)^2]$, where
$N$ is the total number of events analyzed, 
$\phi_i = 
f\,\Delta t_i + \dot{f}\, (\Delta t_i)^2/2$
is the phase, $f$ and $\dot{f}$ 
are the trial ephemeris parameters,
$\Delta t_i$
($i=1,...,N$) is the event arrival time counted from 
an epoch of zero phase.

If the arrival times are uniformly distributed (periodic component
is absent), then the probability density function of $\Z$ is
equal to that of a $\chi^2$ with two degrees of freedom,
with both the mean and the standard deviation equal to 2.
The expected (mean) number
of peaks with $Z_1^2 > \bar{Z}_1^2$ in ${\cal N}$ statistically
independent trials is ${\cal N}\, \exp (-\bar{Z}_1^2/2)$.
This means that if a maximum value
of $\Z$ in an $f,\df$ domain is such that
$\alpha \equiv {\cal N} \exp(-Z_{1,{\rm max}}^2/2) < 1$, periodicity
at the corresponding $f$ and $\df$ can be established at a confidence level
of $C=(1-\alpha)\times 100\%$.

For a
sinusoidal signal with 
frequency $f_0$, frequency derivative $\df_0$, 
and pulsed fraction $p$, 
the mean and the standard deviation of the peak value $\Z(f_0,\dot{f}_0)$
can be estimated as $Np^2/2$ and $(2N)^{1/2}p$, respectively, provided
$Np^2\gg 4$ and $p^2\ll 1$. 
This means that the periodicity can be established at a high
confidence level $C$ only if $N$ and $p$ are so large, and/or ${\cal N}$
so small, that
${\cal N} \exp(-Np^2/4) \le 1-(C/100\%)$. 
If the quantity in the left-hand side of this inequality is not
small enough, the peak corresponding to the true ephemeris parameters
may be comparable to, or even lower than statistical fluctuations of $\Z$
at other values of $f$ and $\df$. In this case, the true parameters
$f_0$ and $\df_0$ cannot be determined reliably  from a single
observation.

The number ${\cal N}$ of statistically independent trials is determined by
the domain of $f$, $\df$ values chosen for the search,
and by the time span $\Ts$ of a given observation.
If $|\df_{\rm max} - \df_{\rm min}| \Ts^2/2 \ll 1$,
then $Z_1^2$ is almost independent of $\df$, and one can use
a one-dimensional (frequency) grid
at a fixed $\df$ from the range chosen.
The number of independent trials 
is ${\cal N} = (f_{\rm max}-f_{\rm min}) \Ts$, and the number
of grid points 
should be greater by a factor 
of $\sim 10$, in order
not to miss the peak. In the opposite case, 
$|\df_{\rm max} - \df_{\rm min}| \Ts^2/2 \gapr 1$, 
one has to use a two-dimensional $f,\df$ grid, 
with
${\cal N}=(f_{\rm max}-f_{\rm min}) |\df_{\rm max} - \df_{\rm min}| \Ts^3/2$
and the number of the grid points of
$\sim 10^2 {\cal N}$.

A reasonable $f,\df$ domain for a young, 
isolated NS is $0.01\lapr f \lapr 100$~Hz,
$-1\times 10^{-10} \lapr \df \le 0$~Hz~s$^{-1}$. The 3~ms accuracy of the 
SCC-UTC conversion hampers detection of high-frequency modulation,
so that we choose  
a conservative upper limit, $f\le 30$~Hz.
For the $f,\df$ domain chosen, $\Z$ is practically independent
of $\df$ for the PSPC observation 
($|\df|_{\rm max} \Ts^2/2=
0.14$), contrary to the HRI observation 
($|\df|_{\rm max} \Ts^2/2=17$), and the numbers of independent trials
are ${\cal N}=1.6\times 10^6$
and $2.9\times 10^8$, respectively.
Because of the enormous number
of the two-dimensional
grid points  needed for the blind periodicity search in the HRI
data,
we have to start from investigating $\Z$ for the PSPC events,
the first step of our timing analysis.
We put $\df = 0$ and calculated $\Z(f)$ at equally spaced 
$2\times 10^7$ frequencies in the range $0.01$--$30$~Hz.
The result is not encouraging --- 
the highest peak, $\Z=32.0$, is not much higher than several other
peaks ($\Z = 30.5$, 29.8, 29.1, 27.8), and the probability that $\Z>32$
for uniformly distributed arrival times is as large as 0.18.
Although
any one of these
peaks could be caused by real pulsations,
with an expected pulsed fraction
$p=20.9 (\Z/30)^{1/2}\%$ for a sine-like pulse profile,  
the corresponding confidence level
would be too low (e.g., 82\%, or 1.3$\sigma$, for $\Z=32$) to claim
the periodicity is detected. 
Thus, we have to conclude that only an upper limit on the pulsed
fraction, $\approx 20\%$, can be established from the PSPC data
alone, in accordance with PBW96.

However, with one more set of timing data available, we can 
examine whether the separate PSPC $\Z$ peaks 
correspond to HRI $\Z$ peaks at reasonably shifted frequencies.
If a PSPC peak at $f=\fp$ is caused by a periodic signal, then
$\Z$ calculated with the HRI arrival times should have a peak at 
a frequency $\fh =\fp +\df T$, where 
$T=1661\fd 00003469$ 
is the time in the TDB scale
between the epochs (start times) 
of the PSPC and HRI observations. 
If such a peak is not
found for reasonable values of $\df$, we have to conclude that
the PSPC peak under examination is most likely a statistical
fluctuation. Thus, the second step of our analysis is 
a conditional search for periodicity in narrow strips
of the $\fh$,$\df$ plane.
The width of the strip along the $\fh$ axis,
$\Delta \fh = 2/T_{\rm span}^{\rm PSPC} \simeq 4\times 10^{-5}$~Hz,
is determined by the uncertainty of $\fp$ due to the finiteness of the
time span. The size of the strip along the $\df$ axis is merely
$|\df|_{\rm max}=1\times 10^{-10}$~Hz s$^{-1}$. The number of
statistically independent trials for each of the strips
can be estimated as ${\cal N} = \Delta\fh \,
|\df|_{\rm max} (T_{\rm span}^{\rm HRI})^3/2 \simeq 370$.

We examined the ten highest PSPC peaks 
and found no corresponding HRI peaks of a high
significance for nine of them 
--- heights of the HRI peaks do not exceed $\Z =16.5$,
which corresponds to confidence levels $C<90.4\%$ ($<1.7\sigma$).
Only for one PSPC peak, $\Z =30.5$ at $\fp = 13.2875651(8)$~Hz, 
a highly significant ($C>99.95\%$, or $>3.5\sigma$)
HRI counterpart was found, $\Z=27.3$ at
$\fh=13.28378836(7)$~Hz 
 and $\df=-2.63(2)\times 10^{-11}$~Hz~s$^{-1}$.  
The digits in parentheses indicate 
uncertainties,
$q/\Ts$ and $2q/\Ts^2$, of the last digits of $f$ and $\df$,
where $q\simeq (3~{\rm ms})\, f_0\simeq 0.04$ is a phase resolution
(cf. Mattox et al.~1996).
The values of the PSPC and HRI ephemeris parameters are for 
the (TDB) epochs 
${\rm MJD}~48361\fd 50067344$ 
and 
${\rm MJD}~50022\fd 50070813$, respectively. The period and its
derivative, for the latter epoch, 
are $P=75.2797300(4)$~ms, 
$\dot{P}=1.49(1)\times 10^{-13}$~s~s$^{-1}$.
To verify that the PSPC and HRI $\Z$ peaks are indeed associated
with \ns, 
we applied the $\Z$ test to
comparable numbers of off-source events in the PSPC and
HRI observations and obtained $\Z$ values close to the mean value
expected for a nonpulsed signal.
\begin{figure}
\vskip -2truein
%\plotone{fig1.ps}
\plotfiddle{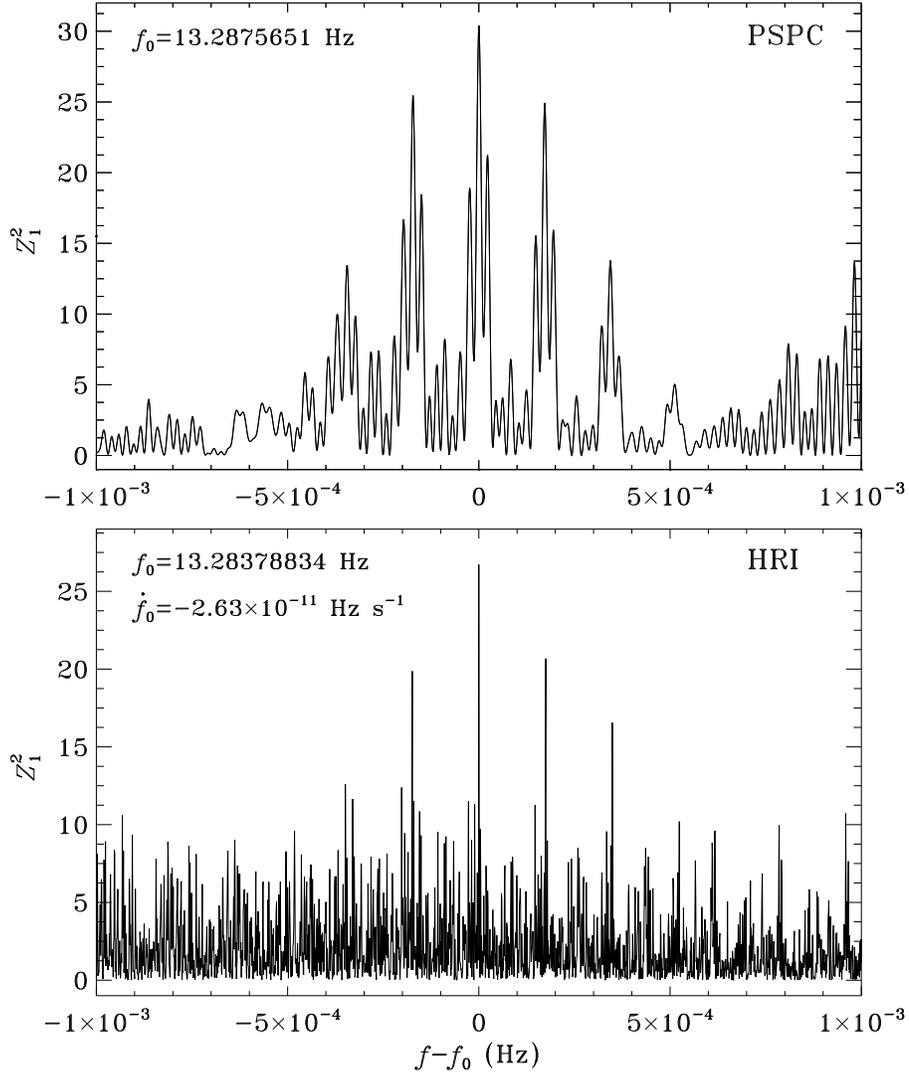}{6truein}{0}{80}{80}{-250}{-200}
\vskip 1truein
\caption{Power spectra around the pulsar frequency $f_0$ for
the PSPC (1991 April) and HRI (1995 November)
observations, at the frequency derivative
$\df =-2.63\times 10^{-11}$~Hz~s$^{-1}$.
}
\end{figure}
\begin{figure}
\vskip -2truein
%\plotone{fig2.ps}
\plotfiddle{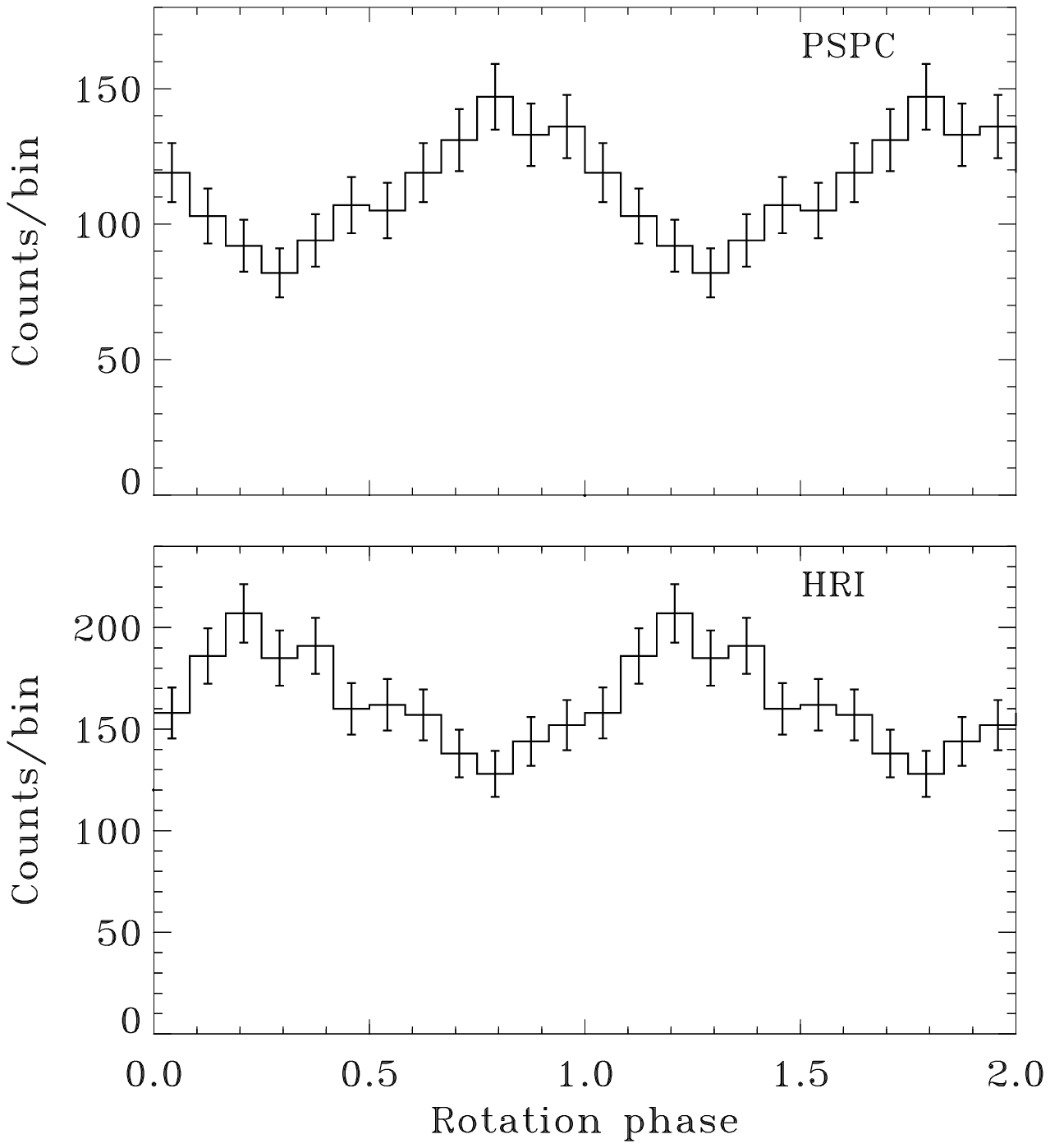}{6truein}{0}{80}{80}{-250}{-200}
%\vskip 0.5truein
\caption{Folded light curves for the PSPC and HRI data.
}
\end{figure}
Figure~1 shows the frequency dependences $\Z(f)$ at fixed $\df=\df_0$
for the PSPC and HRI data. The main (highest)
peaks at $f=f_0$ are accompanied by a few
side peaks at $f=f_0\pm 1.74\times 10^{-4} k$~Hz ($k=1, 2, \ldots$),
caused by exposure gaps associated with the \rosat
revolution around Earth with the period
of 96.0~min.
Both the main and the side peaks have an additional fine structure
with characteristic
``periods'' $1/\Ts^{\rm PSPC}=1.9\times 10^{-5}$~Hz and
$1/\Ts^{\rm HRI}=1.7\times 10^{-6}$~Hz.
Similar coherent structures can be also seen in
the dependence of the HRI $\Z$ on $\df$ at a fixed $f$, 
with a shortest ``period''
$2/(\Ts^{\rm HRI})^2 = 5.9\times 10^{-12}$~Hz~s$^{-1}$.

Figure~2 shows the pulse profiles extracted 
by folding the arrival times with the 
ephemeris parameters obtained 
for the PSPC and HRI data sets. Each of the curves is drawn for 
the corresponding 
epoch, which causes the apparent 
phase shift. This shift could be eliminated by 
combining the PSPC and HRI events in one set and determining
the ephemeris parameters with a much higher accuracy,
but the total time span, $\Ts
\simeq 1.441\times 10^8$~s, is so long that this cannot be
done unequivocally.
We verified that there is no statistically significant
difference between the shapes of the PSPC and HRI pulses. 
The light curves binned in 12 phase
bins have broad maxima consistent with a thermal origin of the
X-ray flux. 
The powers of higher harmonics 
are small in comparison with $\Z$
(3.2 vs. 30.5 and 1.0 vs. 27.3),
which demonstrates the pulse profile shape is close to a sinusoid
and justifies {\it a posteriori} our choice of $\Z$ as a test statistic.
To estimate the pulsed fractions, we applied the
bootstrap method proposed by Swanepoel, de Beer \&
Loots (1996) and obtained
$p=22.5\pm 4.2\%$ and $20.1\pm 3.5\%$
for the PSPC and HRI light curves,
respectively. 
We also calculated the PSPC pulse profiles at lower and higher
photon energies (0.1--1.1 and 1.1--2.4 keV)
and found no statistically significant phase shift or difference of
the pulse shapes.
%%%%%%%%%%%%%%%%%%%%%%%%%%%%%%%%%%%%%%%%%%%%%%%%%%%%%%%%%%%%%%%%%%%%%
%%%%%%%%%%%%%%%%%%%%%%%%%%%%%%%%%%%%%%%%%%%%%%%%%%%%%%%%%%%%%%%%%%%%%
\section{Discussion}
With the period and its derivative evaluated, it is straightforward
to estimate the pulsar's age, magnetic field, and rotational energy loss.
Assuming a pulsar spin-down equation $\df= -K f^n$, where $n$ is the
braking index and $K={\rm const}$, the characteristic
(dynamical) pulsar's age equals $\tau=-f/[(n-1)\df]=16.0/(n-1)$~kyr, 
provided that the rotational rate at the NS birth was much greater
than its current value, $f_{\rm in}\gg f$.
For $n=3$, corresponding to a simple magnetic-dipole braking, 
we obtain $\tau=8.0$~kyr.
If we accept that the
true pulsar age coincides with the the kinematic age of the SNR, 
$3.7\pm 0.4$~kyr,
then the braking index should be $n=5.3\pm 0.5$.
To match the pulsar's age and the kinematic SNR age 
at $2.2<n<2.9$, observed for the youngest radio pulsars
PSR B0531+21, B0540--69, B1509--58, 
$f_{\rm in}$ should 
be within a 16.8--18.9~Hz range ($P_{\rm in}=53$--59~ms).
A remarkably similar situation with the
SNR/pulsar age and the braking index is found for PSR J1617--5055 near the
SNR RCW~103 
(Torii et al.~1998; Kaspi et al.~1998).
The conventional 
``magnetic field'' of the pulsar, derived from the standard formula
$B=1.0\times 10^{12}(P \dot{P}_{-15})^{1/2}$~G 
(e.~g., Manchester \& Taylor 1977),
equals $3.4\times 10^{12}$~G 
for 
a standard moment of inertia ,
$I=10^{45}$~g~cm$^2$, and a NS radius of 10~km.
The rotational energy loss can be estimated as $\dot{E}=-4\pi^2 I \df f=
1.4\times 10^{37}$~erg~s$^{-1}$. The parameters inferred for \ns are
close to those of the Vela pulsar, $P=89.3$~ms, $\dot{P}=1.25\times
10^{-13}$, $P/(2\dot{P})=11$~kyr, $B=3.4\times 10^{12}$~G,
$\dot{E}=0.7\times 10^{37}$~erg~s$^{-1}$.

These results indicate that RX~J0822--4300 may be a regular radio
pulsar whose radio-quiet nature can be explained by an
unfavorable orientation of the pulsar beam. 
Contrary to the
radio flux, the X-rays observed from RX~J0822--4300 
originate from the whole NS surface (Zavlin et al.~1998b), their
pulsation with $p\simeq 20\%$ may be due to anisotropy of the surface
temperature distribution and magnetized atmosphere radiation.
However, it is expected that the lower is the frequency, the broader
is the radio beam. Therefore, deep radio observations at low
frequencies could detect 
RX~J0822--4300,
as it has happened recently with
Geminga (Kuzmin \& Losovsky 1997; Malofeev \& Malov 1997).

Such a young, energetic pulsar is expected to power 
a pulsar wind nebula, either in the form of a plerion or a bow shock.
The \rosat images show two ``blobs'',
$\approx 2'$ north and $\approx 1'$ south of the NS,
with luminosities $L_x\sim (1-2)\times 10^{34}d_2^2$~erg~s$^{-1}$
in the \rosat range; the blobs are apparently
connected to the NS with strips of lower brightness.
In addition, the \asca\ GIS image at higher energies,
$E>3$~keV, reveals a structure resembling a bow-shock nebula,
about $5'-6'$ east of RX~J0822--4300. 
Although these structures might
be manifestations of the pulsar activity, one still cannot firmly
exclude that they belong to the SNR and are not physically
associated with the pulsar.
Future X-ray and radio observations of \ns and its environment
are needed 
to elucidate the nature of these structures.
%%%%%%%%%%%%%%%%%%%%%%%%%%%%%%%%%%%%%%%%%%%%%%%%%%%%%%%%%%%%%%%%%%%%%%%%%%%%
%%%%%%%%%%%%%%%%%%%%%%%%%%%%%%%%%%%%%%%%%%%%%%%%%%%%%%%%%%%%%%%%%%%%%%%%%%%
\acknowledgments 
It is our pleasure to thank Rob Petre, the referee, for
very valuable comments.
%%%
The data preparation for the timing analysis was done with the $EXSAS$
software developed at MPE.
%%%
This work was partially supported through 
NASA grants NAG5-6907 and NAG5-7017.
%%%%%%%%%%%%%%%%%%%%%%%%%%%%%%%%%%%%%%%%%%%%%%%%%%%%%%%%%%%%%%%%%%%%%%%%%%%

%%%%%%%%%%%%%%%%%%%%%%%%%%%%%%%%%%%%%%%%%

\begin{references}

\reference{}
Anderson, S.~B., Cordova, F.~A., Pavlov, G.~G., Robinson, C.~R.,
\& Thomson, R.~J. 1993, ApJ, 414, 867

\reference{}
Brazier, K.~T.S., \& Johnston, S. 1998, MNRAS, submitted

\reference{}
Buccheri, R., et.~al. 1983, A\&A, 128, 245

\reference{}
De Jager, C.~O. 1994, ApJ, 436, 239

\reference{}
Greenstein, G., \& Hartke, G.~J. 1983, ApJ, 271, 283

\reference{}
Kaspi, V.~M., 
et~al. 1998, ApJ, 503, L161

\reference{}
Kuzmin, A.~D., \& Losovsky, B.~Y. 1997, Astron. Let., 23, 323

\reference{}
Malofeev, V.~M., Malov O.~I. 1997, Nature, 389, 697

\reference{}
Manchester, R.~N., \& Taylor, J.~H. 1977, Pulsars (San Francisco: Freeman)

\reference{}
Mattox, J.~R., Halpern, J.~P., \& Caraveo, P.~A. 1996,
A\&AS, 120, 77 

\reference{}
Petre, R., Canizares, C.~R., Kriss, G.~A., \& Winkler, P.~F.
1982, ApJ, 440,706

\reference{}
Petre, R., Becker, C.~M., \& Winkler, P.~F. 1996, 465, L43 (PBW96)

\reference{}
Reynoso, E.~M., Dubner, G.~M., Goss, W.~M., \& Arnal, E.~M. 1995,
AJ, 110, 318

\reference{}
Shibanov, Yu.~A., Pavlov. G.~G., Zavlin, V.~E., Qin, L., \& Tsuruta, S.
1995, in The 17th Texas Symposium on Relativistic
Astrophysics and Cosmology, eds. H. B\"ohringer, G.~E. Morfill \&
J.~E. Tr\"umper (NY: NY Academy of Science), 759, p.~291

\reference{}
Swanepoel, J.~W.~H., de Beer, C.~F., Loots, H. 1996, ApJ, 467, 261

\reference{}
Torii, K., et~al. 1998, ApJ, 494, L207

\reference{}
Vasisht, G., \& Gotthelf, E.~F. 1997, ApJ, 486, L29

\reference{}
Winkler, P.~F., Tuttle, J,~H., Kirshner, R.~P., \&
Irwin, M.~J. 1988, in Supernova
Remnants and the Interstellar Medium, eds. R.~S. Roger \& T.~L.
Landecker (Cambridge: Cambridge Univ. Press), p.~65

\reference{}
Zavlin, V.~E., Pavlov, G.~G., \& Tr\"umper, J. 1998a, A\&A, 331, 821

\reference{}
Zavlin, V.~E., Pavlov, G.~G., \& Tr\"umper, J. 1998b, ApJ, in preparation

\end{references}
\end{document}